\documentclass[debug,overfull]{epl}

\usepackage{epsfig}
\usepackage{amssymb}
\usepackage{bm}

\newcommand{\vex}{\vec{x}}

\newcommand{\vecc}{\vec{c}}

\newcommand{\be}{\begin{equation}}
\newcommand{\ee}{\end{equation}}

\title{A note on the Lattice Boltzmann Method beyond the Chapman-Enskog limits}
\shorttitle{Lattice BGK beyond Chapman-Enskog}

\author{M. Sbragaglia$^1$, S. Succi$^2$}

 \institute{$^1$ Dipartimento di Fisica, Universit\`a ``Tor
  Vergata'', and INFN,\\Via della Ricerca Scientifica 1, I-00133 Roma,
  Italy.\\ $^2$ CNR, IAC, Viale del Policlinico 137, I-00161 Roma,
  Italy.} 


\begin{document}

\maketitle

\begin{abstract} 
A non-perturbative analysis of the Bhatnagar-Gross-Krook (BGK) 
model kinetic equation for finite values of the Knudsen
number is presented.
This analysis indicates why discrete kinetic versions of the BGK 
equation, and notably the Lattice Boltzmann method, can provide 
semi-quantitative results also in the non-hydrodynamic, 
finite-Knudsen regime, up to $Kn\sim {\cal O}(1)$. 
This may help the interpretation of recent Lattice Boltzmann simulations 
of microflows, which show satisfactory agreement with continuum kinetic
theory in the moderate-Knudsen regime.
\vskip 0.2cm 
\end{abstract}

In the last decade, the lattice Boltzmann Method (LB) has developed 
into a very flexible and effective numerical method for the simulation 
of a large variety of 
complex flows, mostly in the macroscopic domain \cite{LBEBVS,wolf,saurobook}. 
Fueled by relentless progress in micro, nano and bio-sciences,
the recent years have witnessed a growing interest in exploring 
the possibility to enrich LB 
in the direction of describing micro-structured flows 
\cite{MICROho,MICROkarni,MICROtabebook}.
\\The LB method is based on a  stylized stream-and-collide microscopic dynamics 
of fictitious particles, located on the nodes
of discrete lattices and interacting according to 
local collision rules that drive the system towards a local 
equilibrium \cite{saurobook,wolf}. 
Mathematically:
$$
f_{i}(\vex+\vecc_{i}\Delta t,t+\Delta t)-f_{i}(\vex,t)=-\omega \Delta t \left[ f_{i}(\vex,t)-f^{(eq)}_{i}(\vex,t) \right]
$$
where $f_{i}(\vex,t)$ is the probability to find a particle at position $\vex$ and time $t$, moving along the lattice direction defined by the 
discrete speeds $\vecc_{i}$ ($i=1,...,b$). 
The second term at the rhs of the above equations denotes 
relaxation towards a local equilibrium, the lattice 
analogue of a Maxwellian distribution in continuum kinetic theory: 
$$
f^{eq}_{i}(\vex,t)=n w_i (1+\beta \vecc_i \cdot \vec{u} + \frac{\beta^2}{2}[(\vecc_i \cdot \vec{u})^2
-u^2].
$$
In the above, $n(\vex,t)$ is the fluid density, $\vec{u}$ the flow speed, $\beta=1/c_s^2$ is the inverse square 
speed of sound, and $w_i$ a discrete set of weights normalized to unity.
Finally, $\omega$, indicates a typical time-scale relaxation (frequency relaxation) to local 
equilibrium, and governs the kinematic viscosity of the LB macroscopic fluid \cite{wolf,LBGK}. 
More general versions accounting for multiple-time relaxation \cite{HSB,HJ,MTR} could be investigated
but the simpler single-time (BGK) relaxation form 
will be sufficient for our present purposes.

By taking suitable averages over molecular speeds, it can be shown  that the resulting macroscopic quantities obey the Navier-Stokes equation of continuum mechanics. The fluid and current density are given by a 
linear superposition of the discrete distributions $f_i$:
\be\label{macro}
n(\vex,t)=\sum_{i=1}^{b} f_{i}(\vex,t), \hspace{.2in} \vec{J}(\vex,t)=\sum_{i=1}^{b} \vecc_{i}f_{i}(\vex,t).
\ee
In order for those quantities to satisfy the exact hydrodynamic 
equations, it is required that the macroscopic fields do not vary appreciably 
on the scale of the mean free path $\lambda$ and the lattice spacing $\Delta x$. 
Based on the consolidated Chapman-Enskog background \cite{CHAPE}, one
might be led to conclude that the range of applicability of 
LB methods is bounded by the domain of validity of the Chapman-Enskog method \cite{CER}, the implication being that LB can only be used for strictly hydrodynamic purposes. 
Yet, such a restrictive stance is challenged by a number of recent numerical 
simulations \cite{SLIP,MICROLBEnie,MICROLBEhe,MICROLBEli,MICROLBElim,EMERSON,AK,AKO,TOSCHI,SBRAGA}  
which clearly show that, by using appropriate boundary 
conditions, LB can reproduce some salient features of flows 
beyond the hydrodynamic regime, such as the onset of slip 
flow at finite-Knudsen numbers. 

In this Letter, we wish to propose a theoretical explanation for this 
rather unexpected validity of LB in the beyond-Chapman-Enskog region. 
Our analysis is based on a non-perturbative solution of the 
lattice BGK (LBGK) equation, which overcomes the restrictions imposed to the 
standard Chapman-Enskog multiple scale analysis and extensions thereof
\cite{SONE,INA}.

Our analysis is confined to the bulk region of the flow, giving for 
granted that the use of proper boundary conditions is crucial to 
obtain correct results in actual LB simulations of microflows
\cite{NOBLE}.\\ 

Let us refer 
for simplicity to the $1d$-continuum Boltzmann equation, written in BGK \cite{BGK} form :
$$\partial_{t}f(x,v,t)+v \partial_{x} f(x,v,t)=-\omega  \left[ f(x,v,t)-f^{(eq)}(x,t) \right ]$$
where the local equilibrium $f^{(eq)}(x,t)$ depends on $x$ and $t$ via the local velocity and density fields. 
Let us now consider the exact solution as 
provided, for each given velocity $v$, by an integration along the particle trajectory, from $t$ to $t+\Delta t$:
\begin{eqnarray}\label{green}
f(x+v \Delta t,t+\Delta t)=e^{-\omega \Delta t}f(x,t)+ \nonumber \\ + \omega e^{-\omega \Delta t} \displaystyle\int_{0}^{\Delta t} e^{s \omega} f^{(eq)}(x+vs,t+s) ds
\end{eqnarray}
being $\Delta t$ a generic time increment corresponding to the 
lattice time step. 
The above expression is exact, but purely formal, until one specifies a 
concrete procedure to compute the integral at the right hand side. 
This exhibits a quadratic functional dependence on $f(x+vs,t+s)$, through 
the local equilibrium $f^{eq}$. 
However, one can formally expand the integrand as:
\be\label{series}
f^{(eq)}(x+vs,t+s)=\sum_{n=0}^{\infty} \frac{s^{n} D^{n}}{n!}  f^{(eq)}(x,t)
\ee
where $D \equiv \partial_{t}+v \partial_{x}$ denotes the streaming operator. 
By inserting this expansion into (\ref{green}), we can formally solve the integral exactly,
and obtain:
\be
f_{t+\Delta t}=e^{-\omega \Delta t} f_{t}+\frac{e^{D\Delta t}-e^{-\omega \Delta t}}{1+D/\omega} f^{(eq)}_{t} 
\ee
where we have used the short-hand notation $f_{t+s}=f(x+vs,t+s)$. 
This expression invites a number of useful considerations. 
Let us first recast it in the symbolic propagator form:
\be\label{PPP}
f_{t+\Delta t}=P(\omega \Delta t) f_{t}+ Q(\omega \Delta t,D\Delta t) f^{(eq)}_{t}
\ee
where
\be
P(\omega \Delta t)=e^{-\omega \Delta t}
\ee
is the propagator from time $t$ to time $t+\Delta t$, and
\be
Q(\omega \Delta t,D\Delta t)=\frac{e^{D\Delta t}-e^{-\omega \Delta t}}{1+D/\omega}
\ee
co-propagates the influence of the equilibrium at time $t$ on the solution at time
$t+\Delta t$.\\
By neglecting the equilibrium variations on a scale $\Delta t$, {\it i.e.}:
\be\label{stickedtime}
f^{(eq)}_{t+t^{\prime}} \simeq f^{(eq)}_{t} \hspace{.3in} 0 \le t^{\prime} \le \Delta t
\ee
the propagator $e^{D \Delta t} $ acting on $f^{(eq)}$ is such that 
\be\label{sticked}
e^{D \Delta t} f^{(eq)}_{t} = 0.
\ee  
This leads to a much simplified version of the integral equation:
\be
f_{t+\Delta t}=e^{-\omega \Delta t} f_{t}+(1-e^{-\omega \Delta t}) f^{(eq)}_{t}
\ee
or, equivalently:
\be\label{PPP2}
f_{t+\Delta t}=P(\omega \Delta t) f_{t}+ Q_{LB}(\omega \Delta t) f^{(eq)}_{t}
\ee
where we have defined the LB co-propagator as:
\be
Q_{LB}(\omega \Delta t) \equiv (1-e^{-\omega \Delta t}).
\ee
The stick-to-equilibrium approximation (\ref{sticked}) is tantamount 
to retaining only the zero-th order term in the Taylor 
expansion (\ref{series}), and consequently it is expected to 
work only for small values of $\Delta t$ in the integral of (\ref{green}). 

It is interesting to note that, already at this zero-th order level, and
without any further assumption on $\omega$, it is possible to identify
a fully explicit discrete Boltzmann equation in BGK form.
This reads:
\be\label{LBGK}
f_{t+\Delta t}-f_{t}=-\omega_{LB} (f_{t}-f^{(eq)}_{t})
\ee
with relaxation frequency $\omega_{LB}=(1-e^{-\omega \Delta t})$.\\
It is now instructive to analyze the two extreme limits $\omega \Delta t \gg 1$ and $\omega \Delta t \ll 1$, in the exact solution (\ref{PPP}) first, and then in the discrete Lattice BGK solution (\ref{PPP2}). 
Let us begin with the former. 
Provided that the kinetic operator $D\Delta t$ remains 
bounded, we can identify the {\it enslaving} limit 
($\omega \Delta t \gg 1$), characterized by:
$$
P(\omega \Delta t)=e^{-\omega \Delta t} \longrightarrow 0
$$
$$
Q(\omega \Delta t,D\Delta t)=\frac{e^{D\Delta t}-e^{-\omega \Delta t}}{1+D/\omega} \longrightarrow e^{D\Delta t}. 
$$
This is equivalent to enslave the solution to the 
local equilibrium, {\it i.e.} $f_{t+\Delta t}=f^{(eq)}_{t+\Delta t} $. 
The opposite situation, ($\omega \Delta t \ll 1$), represents the 
{\it free-molecular} limit, in which one has:
$$
P(\omega \Delta t)=e^{-\omega \Delta t} \longrightarrow 1
$$
$$
Q(\omega \Delta t,D\Delta t)=\frac{e^{D\Delta t}-e^{-\omega \Delta t}}{1+D/\omega} \longrightarrow 0 
$$
corresponding to a free-flow (collisionless) solution $f_{t+\Delta t}=f_{t}$. 

A natural question arises on the nature of the same limits in the Lattice BGK equation (\ref{PPP2})? 
The effect on the streaming propagator is the same, since 
the streaming term is integrated {\it exactly} in the lattice version. 
It remains to inspect the behaviour of $Q_{LB}(\omega \Delta t)$ in the two aforementioned limits. 
The enslaving limit ($\omega \Delta t \gg 1$)  yields:
$$
Q_{LB}(\omega \Delta t)=1-e^{-\omega \Delta t} \longrightarrow 1 \hspace{.2in} (\omega_{LB} \sim {\cal O}(1))
$$
that is $f_{t+\Delta t}=f^{(eq)}_{t}$. 
In view of the relation (\ref{stickedtime}),
this is equivalent to state $f_{t+\Delta t}=f^{(eq)}_{t+\Delta t}$.
The opposite situation ($\omega \Delta t \ll 1$) yields:  
$$
Q_{LB}(\omega \Delta t)=1-e^{-\omega \Delta t} \longrightarrow 0 \hspace{.2in} (\omega_{LB} \rightarrow 0). 
$$
Thus, the two limits, {\it full-enslaving} and {\it free-flow}, are 
recovered by the LBGK, provided that the relaxation frequency is 
turned from bare $\omega$ to $\omega_{LB}$. 
The question remains: what happens inbetween? 
A restrictive tenet is that Lattice BGK cannot work properly because 
the intermediate regime involves all-order tensors, through 
the powers $D^{n}$, which cannot be reproduced correctly in 
the discrete lattice because of lack of symmetry. 
On a more optimistic vein, one could counter-argue that since {\it both}
extreme limits are correctly recovered, there might be hope that
even inbetween the LB method could continue to provide a reasonable
agreement with continuum kinetic theory. 

In order to analyze this point, we introduce the 
{\it fine-scale} Knudsen number $Kn$, defined as:
\be
Kn=\lambda/\delta
\ee
where $\lambda=v/\omega$ is the kinetic mean-free path and 
$\delta$ is the smallest macroscopic length. 

Let us consider the most critical situation 
$\delta=\Delta x=v \Delta t$, {\it i.e.} the macroscopic fields 
show appreciable variation on the scale of a single lattice site. 
Under these specific conditions, we obtain:
\be
\omega \Delta t = 1/Kn
\ee
from which it follows that: 
\be\label{master1}
P(\omega \Delta t)=P(Kn)=e^{-1/Kn} 
\ee
\be\label{master2}
Q(\omega \Delta t,D\Delta t)=Q(Kn,D\Delta t)=\frac{e^{D\Delta t}-e^{-1/Kn}}{1+Kn D\Delta t} 
\ee
\be\label{master3}
Q_{LB}(\omega \Delta t,D\Delta t)=Q_{LB}(Kn,D\Delta t)=1-e^{-1/Kn}.
\ee

\begin{figure}[h]
\centerline{\includegraphics{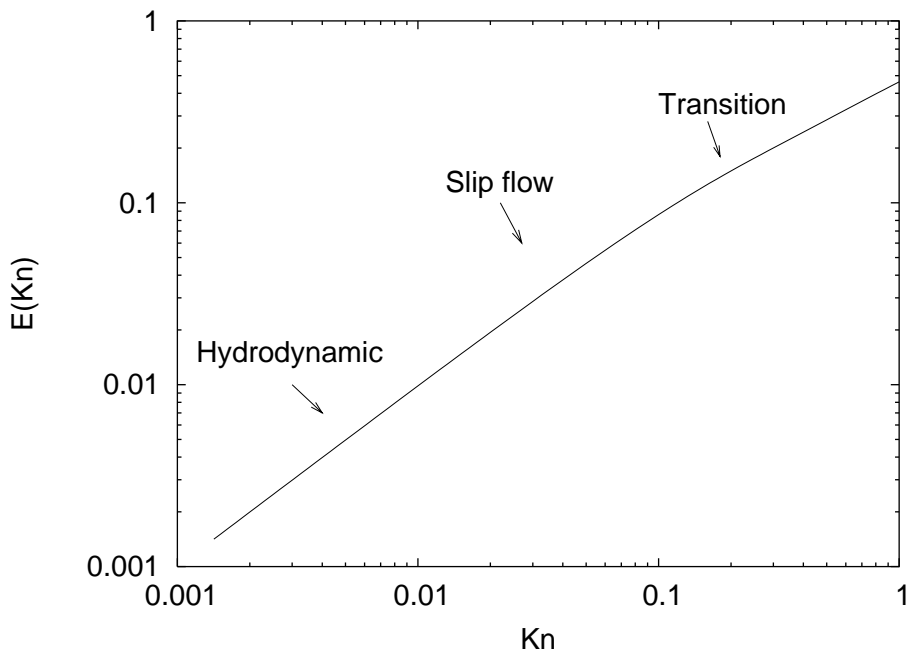}}
\caption{Plot of the percentage error $E$, as defined in (\ref{perc}), as a function of the Knudsen number up to the transition regime. Note that the error between the exact Boltzmann solution and the discrete form reaches a maximum of about ten percent at the top of the slip-flow regime ($Kn \sim 0.1$).}
\label{fig1}
\end{figure}

Our Knudsen number-dependent approach is consistent with the fact that the full enslaving limit is recovered at $Kn \rightarrow 0$, which is equivalent to use a LBGK approach (\ref{LBGK}) with $\omega_{LB} \Delta t=(1-e^{-\omega \Delta t}) \sim {\cal O}(1)$. Starting from this Chapman-Enskog limit ($Kn \rightarrow 0$), it is then interesting to see whether the LB approach can be extended to the {\it slip flow} and {\it transition} regimes  (up to $Kn \sim {\cal O}(1)$). 
In these regimes, the macroscopic fields fluctuate, in the most pessimistic case, on the same scale of the 
kinetic fields. Therefore, we may consistently assume 
that $Df/f \sim v/\delta$, and make the identification: 
\be\label{DKn} 
D \Delta t \sim Kn.
\ee
This is the zero-th order approximation relating $D \Delta t$ to $Kn$. 
More rigorously, one should consider a full series in $Kn$. 
However, to any practical purpose, this series can be truncated at the ${\cal O}(Kn)$ 
without hampering the results up to $Kn \sim {\cal O}(1)$. 
In this approximation, the propagators become:
$$
Q(Kn)=\frac{e^{Kn}-e^{-1/Kn}}{1+Kn^{2}} 
$$
$$
Q_{LB}(Kn)=1-e^{-1/Kn}
$$
where $Q(Kn)$ and $Q_{LB}(Kn)$ 
represent the exact physical co-propagator effects and 
the lattice ones, in the range $Kn<1$, respectively. 
The relative departure:
\be\label{perc}
E(Kn)=\frac{Q(Kn)-Q_{LB}(Kn)}{Q(Kn)}
\ee
is a quantitative measure of the spurious 
lattice-induced effects on the co-propagator.

From figure (\ref{fig1}), where we report $E(Kn)$ as a function of $Kn$, it is 
recognized that, up to $Kn\sim 0.1$, {\it i.e. } in the slip flow regime, the 
lattice Boltzmann approach does not differ from  
the exact solution for more than to within a ten percent. 
Only as $Kn$ exceeds $0.1$, and the transition regime is 
entered, significant error build-up is observed.\\
It is important to emphasize that the above
figures apply to the worst-case scenario, in which macroscopic fields
vary on the scale of a {\it single} lattice spacing 
(an extreme beyond-Chapman-Enskog situation).
One may soften this assumption and assume that macroscopic fields exhibit 
significant variations on scales larger that a single grid spacing 
$\Delta x$, say:
\be
\delta= h \Delta x \hspace{.2in} h=1,2...
\ee
$h=1$ reproducing the previous worst-case scenario. 
This means that the Knudsen number is 
reduced by a factor $h$, $Kn \rightarrow Kn/h$, leading 
to a corresponding reduction in the error $E(Kn)$.
In addition, we observe that $E(Kn)$ measures the error
of the propagator, whereas physical observables result from
the summation of discrete populations $f_i$ over the 
discrete speeds (see equations (\ref{macro})) and consequently $E(kn)$ as 
defined in (\ref{perc}) may well represent a pessimistic bound.
In any event, the point of the present analysis is not to
state the case for the accuracy of LB in the non-hydrodynamic
regime, but only to point out that, even at finite-Knudsen, discreteness effects
remain within fairly tolerable limits, at least for semi-quantitative
purposes. Whether or not LB should be used instead of more
accurate, and much more expensive, methods, such as Direct Simulation
Monte Carlo, remains to be decided on a case-by-case basis.


\vspace{5mm}

Summarizing, a non-perturbative analysis of the 
Boltzmann equation in BGK form, indicates that
the LB method may continue to provide semi-quantitative agreement
beyond the limits of the Chapman-Enskog theoretical framework. 
This could be of interest for the interpretation of Lattice Boltzmann 
simulations in the finite-Knudsen regime, including
kinetic modeling of fluid turbulence \cite{SCI}.
Clearly, in order to realize the bulk properties highlighted by the 
present analysis, proper kinetic boundary conditions, 
\cite{AK,SBRAGA}, must be used in actual LB simulations 
of finite-Knudsen flows. 

\vspace{5mm}

R. Benzi, L. Biferale and F. Toschi are kindly acknowledged for useful
comments and critical reading of this manuscript.


\begin{thebibliography}{}


\bibitem{LBEBVS} Benzi R., Succi S. and Vergassola M., \Review{Phys. Rep.} \Vol{222} \Year{1992} \Page{145}.

\bibitem{wolf} Wolfe-Gladrow D., \Book{Lattice Gas Cellular Automata and Lattice Boltzmann Model} \Editor{Springer Verlag, Berlin} \Year{2000}.

\bibitem{saurobook} Succi S., \Book{The Lattice Boltzmann equation} \Editor{Oxford Univ. Press, Oxford} \Year{2001}.

\bibitem{MICROho} Ho C. M. and Tai Y. C., \Review{Annu. Rev. Fluid Mech.} \Vol{30} \Year{1998} \Page{579}.

\bibitem{MICROkarni} Karniadakis G. and Beskok A., \Book{Microflows} \Editor{Springer Verlag, Berlin} \Year{2002}.

\bibitem{MICROtabebook} Tabeling P., \Book{Introduction a la microfluidique} \Editor{Belin, Paris}  \Year{2003}.

\bibitem{LBGK} Qian Y., d'Humi\`{e}res D. and Lallemand P., \Review{Europhys. Lett.} \Vol{17} \Year{1992} \Page{479} .

\bibitem{HSB} Higuera F., Succi S., Benzi R., \Review{Europhys. Lett.} \Vol{9}  \Year(1989) \Page{345}.

\bibitem{HJ} Higuera F., Jimenez J., \Review{Europhys. Lett.} \Vol{9} \Year{1989}  \Page{663}.

\bibitem{MTR} D' Humi\`{e}res D., \Review{Prog. Astronaut. Aeronaut.} \Vol{159} \Year{1992} \Page{450}.

\bibitem{CHAPE} Chapman S. and Cowling T. G., \Book{The mathematical theory of non uniform gases} \Editor{Cambridge University Press, Cambridge} \Year{1991}.

\bibitem{CER} Cercignani C., \Book{Theory and application of the Boltzmann equation} \Editor{Scottish Academic Press, New York} \Year{1975}.

\bibitem{SLIP} Succi S., \Review{Phys. Rev. Lett.} \Vol{89} \Year{2002} \Page{064502}.

\bibitem{MICROLBEnie} Nie X., Doolen G. and Chen S., \Review{J. Stat. Phys.} \Vol{107} \Year{2002} \Page{29}.

\bibitem{MICROLBEhe} He X., Zou Q., Luo L.S. and Dembo M., \Review{J. Stat. Phys.} \Vol{87} \Year{1997} \Page{115}.

\bibitem{MICROLBEli} Li B. and Kwok D., \Review{Phys. Rev. Lett.} \Vol{90} \Year{2003}  \Page{124502}.

\bibitem{MICROLBElim}Lim C., Shu C., Niu X. and Chew Y., \Review{Phys. Fluids} \Vol{14} \Year{2002} \Page{2299}.

\bibitem{EMERSON} Zhang Y., Qin R., and Emerson D., \Review{Phys. Rev. E}  \Vol{71} \Year{2005} \Page{047702}.

\bibitem{TOSCHI} Toschi F. and Succi S., \Review{Europhys. Lett.} \Vol{69} \Year{2005} \Page{549}.

\bibitem{SBRAGA} Benzi R., Biferale L., Sbragaglia M., Succi S.  and Toschi F., \Review{J. Fluid. Mech.}, in press.

\bibitem{AK} Ansumali S. and  Karlin I., \Review{Phys. Rev. E} \Vol{66} \Year{2002} \Page{026311}.

\bibitem{AKO} Ansumali S., Karlin I. and Oettinger C.,\Review{Europhys. Lett.} \Vol{63} \Year{2003} \Page{798}.

\bibitem{INA} Inamuro T., Yoshino M. and Ogino F., \Review{Phys. of Fluids} \Vol{9} \Year{1997} \Page{3535}.

\bibitem{SONE} Sone Y., \Book{Rarefied gas dynamics}, \Editor{Editrice tecnico scientifica, Pisa} \Vol{2} \Year{1971} \Page{737}.

\bibitem{NOBLE} This point was analyzed in
Holdych D. J. et al., \Review{J. Comp. Phys.} \Vol{193} \Year{2004} \Page{594}.
However the authors did not seem to focus on
finite-Knudsen issues, hence did not consider kinetic boundary conditions.


\bibitem{BGK} Bhatnagar L., Gross E. and Krook M., \Review{Phys. Rev.} \Vol{94} \Year{1954} \Page{511}.


\bibitem{SCI} H. Chen, S. Kandasamy, S. Orszag, R. Shock, S. Succi, V. Yakhot,
\Review{Science} \Vol{301} \Year{2003} \Page{633}. 



\end{thebibliography}
\end{document}